\documentclass[prb,twocolumn,showpacs,amsmath,amssymb,preprintnumbers]{revtex4}
\usepackage{dcolumn}
\usepackage{bm}
\usepackage{graphicx}
\usepackage{color}

\begin{document}

\title{Nonequilibrium low temperature phase in pyrochlore iridate Y$_2$Ir$_2$O$_7$: Possibility of glass-like dynamics}

\author{Harish Kumar}\affiliation{School of Physical Sciences, Jawaharlal Nehru University, New Delhi - 110067, India.}
\author{A. K. Pramanik}\email{akpramanik@mail.jnu.ac.in}\affiliation{School of Physical Sciences, Jawaharlal Nehru University, New Delhi - 110067, India.}

\begin{abstract}
Geometrical frustration and spin-orbit coupling effect together play vital role to influence properties in pyrochlore based iridium oxides. Here we have investigated detailed structural, magnetic, thermodynamic and transport properties of pyrochlore iridate Y$_2$Ir$_2$O$_7$. Magnetization data show onset of magnetic irreversibility around temperature $T_{irr}$ $\sim$ 160 K, however, no sign of long-range type ferromagnetic ordering is observed below $T_{irr}$. Specific heat data show no visible anomaly across $T_{irr}$, and the analysis of data indicate sizable density of states across Fermi level. Temperature dependent x-ray diffraction measurements show no change in structural symmetry down to low temperature. The material, on the other hand, shows significant relaxation and aging behavior similar to glassy dynamics. The electronic charge transport in this highly insulating system is found to follow power law dependence with temperature. The material shows negative magnetoresistance which is explained with quantum interference effect.  
\end{abstract}


\maketitle

\section {Introduction}
Pyrochlores with chemical formula A$_2$B$_2$O$_7$ (A and B generally stands for rare earth and transition metal ions, respectively) have attracted surge of scientific interest over last one or two decades \cite{subram,gardner}. These materials have in-built frustration arising from structural geometry where the A and B cations form  corner-shared tetrahedra. As a result, many interesting physical phenomena are observed. Most notable examples are spin glass \cite{gingras,yoshii}, spin ice \cite{bramwell,fukazawa}, spin liquid \cite{gardner1,nakatsuji}, anomalous Hall effect \cite{taguchi,machida}, superconductivity \cite{hanawa}, Kondo-like behavior \cite{machida}, etc.

In recent times, Ir-based pyrochlores (A$_2$Ir$_2$O$_7$) have been at center place for investigation. In addition to its geometric frustration, these materials exhibit sizable spin-orbit coupling (SOC) effect due to presence of heavy Ir atoms. However, the electron-electron correlation effect ($U$), which is usually prominent in transition metal oxides, turns out relatively small in these systems due to extended character of 5$d$ orbitals of Ir. Nonetheless, the SOC and electronic correlation effect exhibit comparable energy scale, therefore, these Ir-based oxides systems offer an ideal playground where the interesting physics coming out as an interplay between these two effects can be studied with control manner. The theoretical calculation, indeed, has predicted various exotic topological insulating phases in pyrochlore iridates based on this intriguing interplay between SOC and electronic correlation effect \cite{pesin}. Moreover, recent study has shown that physical properties in pyrochlore iridates A$_2$Ir$_2$O$_7$ largely depend on ionic radius of $A^{3+}$ where the material gradually shifts from its insulating/magnetic behavior to metallic/nonmagnetic one with increasing ionic radius \cite{matsuhira}.

Within the family of pyrochlore iridates, the Y$_2$Ir$_2$O$_7$ has special interest as it has nonmagnetic Y$^{3+}$ residing at A-site therefore, the magnetic properties are mostly determined by the contribution from Ir sublattice. In addition, one can exclude the possibility of $f$-$d$ exchange interactions which has been discussed to introduce complicated magnetic interactions in Ir-based pyrochlores \cite{chen}. This is also important considering obvious interconnection between magnetic and electronic properties in these materials. This shows Y$_2$Ir$_2$O$_7$ is an ideal system to study the intercoupling effect of SOC and $U$ while neglecting other associated contributions. This material by nature is insulator where the resistivity increases by couple of orders at low temperature \cite{soda}. The magnetic state is, however, still debated. While the neutron diffraction/scattering measurements reveal no sign of magnetic long-range ordering \cite{shapiro}, the muon spin rotation/relaxation measurement \cite{disseler}, on the other hand, has shown well-defined spontaneous oscillations in muon symmetry indicating long-range ordering. Considering the in-built frustration in pyrochlore lattice, the glass-like behavior could also be a low temperature magnetic state in Y$_2$Ir$_2$O$_7$. Note, that Y-based other pyrochlores i.e., Y$_2$Mo$_2$O$_7$, Y$_2$Ru$_2$O$_7$ have shown spin-glass behavior \cite{gingras,yoshii}. This underlines the fact that there is need for conclusive understanding of magnetic state in Y$_2$Ir$_2$O$_7$ which is further amplified with the fact that recent calculation proposes topological Weyl semimetallic phase in Y$_2$Ir$_2$O$_7$ based on its ground state magnetic character \cite{wan}.

In this work, we present detailed structural, magnetic, thermodynamic and transport properties of pyrochlore iridate  Y$_2$Ir$_2$O$_7$. Magnetic measurements show onset of magnetic irreversibility at temperature $T_{irr}$ $\sim$ 160 K, however, no sharp peak/cusp is observed at this temperature. Analysis of magnetization data shows no spontaneous moment at low temperature. Instead, reasonable magnetic relaxation and aging behavior similar to metamagnetic systems such as, spin-glasse, superparamagnet has been observed. Temperature dependent x-ray diffraction (XRD) measurements show there is no crystallographic phase/symmetry transition down to low temperature. The material is found to be insulating throughout the temperature range where the resistivity follows power law behavior with temperature. The negative magnetoresistance at low temperature can be explained with quantum interference effect.
                             
\section {Experimental Details}
Polycrystalline samples of Y$_2$Ir$_2$O$_7$ have been prepared using standard solid state method. The ingredient powder materials Y$_2$O$_3$ and IrO$_2$ with phase purity $>$ 99.99\% (Sigma-Aldrich) are taken in stoichiometric ratio. The mixture are ground well, subsequently pelletized and heated in air at 1000$^o$C for 96 hours, at 1100$^o$C for 96 hours and at 1160$^o$C for 252 hours with intermediate grinding. The heating and cooling rate has been used 3$^o$C/min. The material has been characterized by powder x-ray diffraction (XRD) using a Rigaku made diffractometer (model: Miniflex600) with CuK$_\alpha$ radiation at room temperature. The XRD data have also been collected at different temperatures in the range of 300 - 20 K using a PANalytical X'Pert powder diffractomer. The low temperature is achieved using a helium close cycle refregerator (CCR) based cold head where proper temperature stabilization is ensured by waiting sufficiently before collecting data. Data have been collected in the range of 2$\theta$ = 10 - 90$^o$ with a step of $\Delta 2\theta$ = 0.033$^o$ and scan rate of 2$^o$/min. The collected XRD data have been analyzed using Reitveld refinement program (FULLPROF) by Young \textit{et al} \cite{young}. The Rietveld analysis shows the material is in single phase except presence of small fraction of unreacted ingredient materials. It can be noted that after certain heat treatment, no further improvement in terms of unreacted ingredients is observed, rather they are found to grow. Magnetically, the Y$_2$O$_3$ is diamagnetic and IrO$_2$ is paramagnetic therefore, these residual impurities are expected to influence the magnetic properties of desired material minimally. Note, that similar impurity phases have also been observed in previous studies \cite{shapiro,disseler,zhu}.

In general, pyrochlore materials adopt cubic crystallographic phase with \textit{Fd$\bar{3}$m} symmetry \cite{subram}. There are two sets of oxygen atoms, therefore pyrochlore structures consist of four nonequivalent crystallographic positions (Table I). The resulting composition can be written as Y$_2$Ir$_2$O$_6$O$'$. The Ir$^{4+}$ cations are six coordinated and they form corner shared IrO$_6$ octahedra with equal length ($d_{Ir-O}$) of Ir-O bonds. The eight coordinated (six O and two O$'$) Y$^{3+}$ cations form distorted cubes where Y-O bonds are equal in length and larger than the Y-O$'$ ones. The only variable position in pyrochlore structures is $x$ position with O (Table I) which determines its structural stability. For instance, $x$ = 0.3125 gives perfect IrO$_6$ octahedra while $x$ = 0.375 results in perfect cubes around Y. The pyrochlore structure can be viewed as interpenetrating two sub-lattices consists of each type of tetrahedra i.e., Y$_4$O$'$ and Ir$_4$$\diamondsuit$, where $\diamondsuit$ implies empty center-site (8a site). As these tetrahedra's are corner shared so the magnetic atoms sitting at vertices of tetrahedra induces frustration. The Rietveld analysis of our XRD data confirms that Y$_2$Ir$_2$O$_7$ crystallizes in cubic phase with \textit{Fd$\bar{3}$m} symmetry (Fig. 1). The unit cell lattice parameter is found to be $a$ = 10.2445 \AA. The bond lengths ($d$) and angles ($<>$) are determined from structural parameters as given in Table II. In IrO$_6$ octahedra, bond angles $<$O-Ir-O$>$ in basal plane are not equal while opposite ones are same in value. The bond lengths and bond angels for present materials (Table II) shows good agreement with the reported values \cite{shapiro}. DC magnetization data have been collected using a vibrating sample magnetometer (PPMS, Quantum Design), specific heat have been measured following semi-adiabatic method and electrical transport properties have been measured using a home-built insert fitted with Oxford superconducting magnet.

\begin{figure}
	\centering
		\includegraphics[width=8cm]{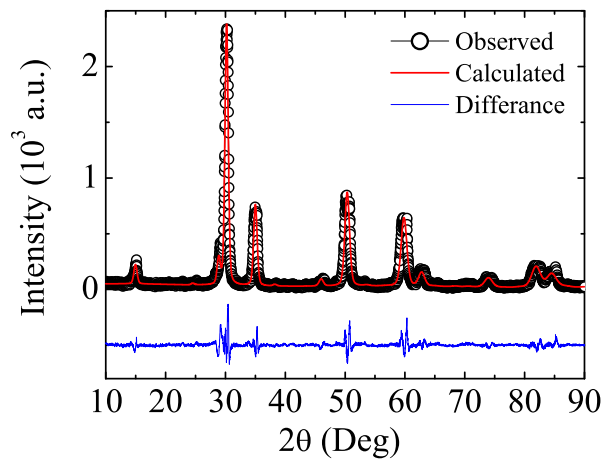}
	\caption{XRD plot for Y$_2$Ir$_2$O$_7$ is shown at room temperature along with Rietveld analysis.}
	\label{fig:Fig1}
\end{figure}

\section{Results and Discussions}
\subsection{Magnetization study}
Magnetization as a function of temperature has been measured following zero field cooled (ZFC) and field cooled (FC) protocol after applying magnetic field of 1 kOe. Fig. 2 shows with decreasing temperature, $M_{ZFC}(T)$ increases, however, there are two observations: around 160 K there is change in slope in data and below around 50 K the magnetization increases very sharply. The corresponding $M_{FC}(T)$ shows identical nature of $M_{ZFC}(T)$ till around 160 K, and below this temperature a bifurcation is observed between both branches of magnetization. The onset of this irreversibility between ZFC and FC magnetization is at temperature $T_{irr}$ $\sim$ 160 K. Some disagreement has earlier been observed about the value of $T_{irr}$. While the majority of studies\cite{kalo,shapiro,taira,zhu} have shown the $T_{irr}$ in the range of 150 - 160 K accompanied by a feeble kink in $M_{ZFC}$ around $T_{irr}$, the another study by Disseler \textit{et al.} \cite{disseler} shows a higher $T_{irr}$ (190 K) and cusp in $M_{ZFC}$ at $T_{irr}$. Our observed $T_{irr}$ $\sim$ 160 K agrees well with the majority of studies. As discussed, the nature of magnetic phase below $T_{irr}$ for this material is debated as both presence as well as absence of long-range ordering has been discussed \cite{shapiro,disseler}. Nonetheless, the bifurcation around $T_{irr}$ in Fig. 2 is suggestive of spin ordering related to long-range or short-range type spin ordering or even spin-glass like freezing. The cusp in magnetic susceptibility, which generally signifies the onset of magnetic ordering, is notably absent in Fig. 2. This similar absence of peak has been observed in other studies as well\cite{kalo,shapiro,taira,zhu}. This qualitatively suggests magnetic frustration is quite significant in this material which perhaps prevent the long-range type magnetic ordering.       

\begin{figure}
	\centering
		\includegraphics[width=8cm]{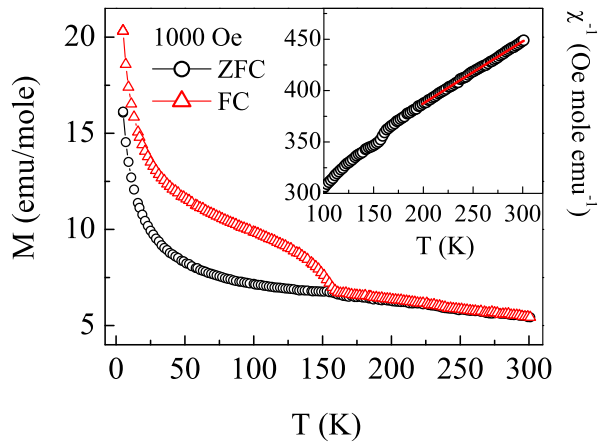}
	\caption{DC magnetization data measured in 1 kOe applied field has been plotted as a function of temperature for Y$_2$Ir$_2$O$_7$. Data are collected following ZFC and FC protocol. Inset shows temperature dependent inverse susceptibility ($\chi^{-1}$ = $(M/H)^{-1}$) deduced from ZFC data for Y$_2$Ir$_2$O$_7$. The solid line is fitting due to modified Curie-Weiss behavior (discussed in text).}
	\label{fig:Fig2}
\end{figure}

In inset of Fig. 2, we show temperature variation of inverse susceptibility ($\chi^{-1}$) which is deduced from ZFC magnetization data $(M_{ZFC}/H)^{-1}$ measured in 10 kOe magnetic field. As evident, above $T_{irr}$, the $\chi^{-1}(T)$ shows a linear behavior over a wide temperature range. The $\chi^{-1}(T)$ has been fitted to modified Curie-Weiss (CW) behavior $\chi$ = $M/H$ = $\chi_0$ + $C/(T - \theta_P)$, where $\chi_0$ is the temperature independent susceptibility, $C$ is the Curie constant and $\theta_P$ is the Curie temperature. The inset of figure shows fitting of data as represented by solid line in temperature regime between 200 - 300 K. The obtained fitted parameters are $\chi_0$ = 1.9 $\times$ 10$^{-4}$ emu/mole, $C$ = 1.4 emu K/mole and $\theta_P$ = -386 K. The sign as well as magnitude of $\theta_P$ imply a reasonably strong non-ferromagnetic type magnetic interaction. Moreover, the obtained $C$ corresponds to effective paramagnetic moment, $\mu_{eff}$ = 3.35 $\mu_B$/f.u. The estimated $\mu_{eff}$ appears higher than the expected value ($g\sqrt{S(S+1)} mu_B$) which turns out to be 1.73 $\mu_B$/f.u in case of spin-only value ($g$ = 2 and $S$ = 1/2). These pyrochlore compounds are inherently frustrated due to their geometrical configuration which often lead to exotic magnetic ground state such as, spin liquid, spin ice, spin glass, etc. The relevant frustration parameter ($f$), which is conventionally defined as ratio $\left|\theta_P\right|$/$T^*$, where $T^*$ is the ordering or spin freezing temperature, has been calculated to be $\sim$ 2.4 taking $T^*$ = 160 K. This $f$ value in present material is significant though much lower than the value ($>$ 100) usually seen for highly frustrated spin-liquid compounds as in Pr$_2$Ir$_2$O$_7$ \cite{nakatsuji}.

\begin{figure}
	\centering
		\includegraphics[width=8.5cm]{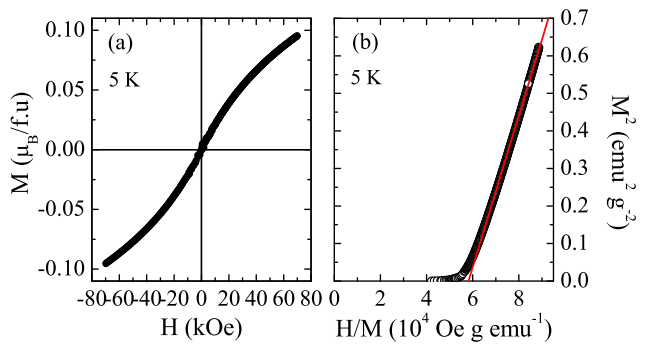}
	\caption{((a) Magnetic field dependent magnetization is shown for Y$_2$Ir$_2$O$_7$ at 5 K. (b) Arrott plot ($M^2$ vs $H/M$) of magnetization data is shown at 5 K.}
	\label{fig:Fig3}
\end{figure}
 
The magnetic field dependent of magnetization data recorded at 5 K in the field range of $\pm$70 kOe are presented at Fig. 3a. The $M(H)$ data do not show linear behavior, and no sign of saturation is observed till 70 kOe. The moment at 70 kOe is found to be 0.095 $\mu_B$/f.u which is much lower than the expected spin-only value ($gS\mu_B$) i.e., 1 $\mu_B$/f.u for S = 1/2. This low value of magnetization is consistent with other studies \cite{shapiro,zhu}. A close observation, however, shows a very small hysteresis in $M(H)$ plot with coercive field $H_c$ $\sim$ 100 Oe and remanent magnetization $M_r$ $\sim$ 4.7 $\times$ 10$^{-4}$ $\mu_B$/f.u. To examine the magnetic state we have analyzed the $M(H)$ data in term of Arrott plot\cite{arrott} ($M^2$ vs $H/M$) as shown in Fig. 3b. The significance of Arrott plot is that if slope is taken at high field regime and yields an intercept on positive $M^2$ axis, this implies presence of spontaneous magnetization or FM ordering. As evident in Fig. 3b, the plot renders intercept at negative $M^2$ axis. This clearly shows absence of spontaneous magnetization viz-a-viz FM ordering in present Y$_2$Ir$_2$O$_7$ material. This finding is important because long-range type antiferromagnet with non-collinear spin ordering (i.e., all-in/all-out type) would exhibit a weak ferromagnetic behavior which will also be manifested in Arrott plot similar to Sr$_2$IrO$_4$ \cite{imtiaz}. Nonetheless, the small hysteresis in $M(H)$ may possibly originate from short-range type FM ordering or freezing of spins at low temperatures. 

Fig. 4 shows temperature dependence of specific heat ($C$) in form of $C/T$ vs $T$ for Y$_2$Ir$_2$O$_7$. The specific heat does not exhibit any observable anomaly around temperature $T_{irr}$. Further, the $C(T)$ data do not show any anomaly or upturn at low temperature which negates the Schottky-like contribution. At low temperature, the $C(T)$ data can be well explained with $C/T$ = $\gamma$ + $\beta T^2$, where $\gamma$ and $\beta$ are the electronic and lattice coefficients of specific heat, respectively \cite{gopal}. Interestingly, the linearity in $C/T$ vs $T^2$ extends to quite high temperature ($\sim$ 17 K). The straight line fitting (inset of Fig. 4) yields $\gamma$ = 32.0(5) mJ mol$^{-1}$ K$^{-2}$ and $\beta$ = 0.592(5) mJ mol$^{-1}$ K$^{-4}$. Using this $\beta$, we have calculated Debye temperature $\theta_D$ (= [(12$\pi^{4}Rn$)/(5$\beta$)]$^{1/3}$, where $R$ is the molar gas constant and $n$ is the number of atoms per formula unit) to be 330.7 K. The $\beta$ and $\theta_D$ is consistent with the reported studies. The $\gamma$, on other hand, is not consistent in literature as both zero\cite{soda,fukazawa1} as well as finite\cite{taira} $\gamma$ is reported which is quite intriguing considering insulating behavior of this material. Using obtained $\gamma$ we have calculated density of states (DOS) at Fermi surface $N(\epsilon_F)$ with following relation,

\begin{figure}
	\centering
		\includegraphics[width=8cm]{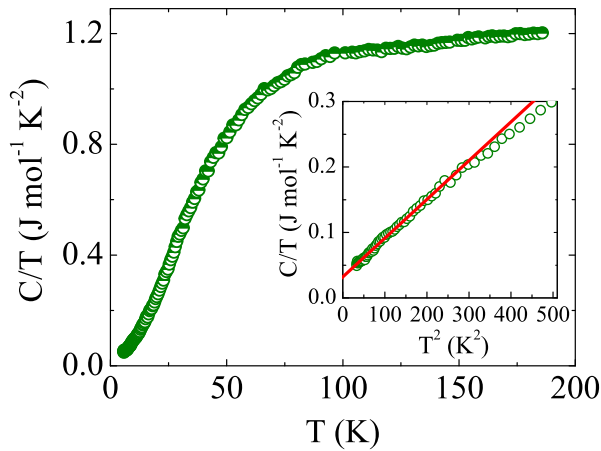}
	\caption{(Temperature dependence of specific heat $C/T$ is shown for Y$_2$Ir$_2$O$_7$. Inset shows $C/T$ vs $T^2$. The straight line is due to linear fitting to $C/T$ = $\gamma$ + $\beta T^2$ (see text).}
	\label{fig:Fig4}
\end{figure}

\begin{eqnarray}
	\gamma = \gamma_0 (1 + \lambda_{e-ph})
\end{eqnarray}

\begin{eqnarray}
	\gamma_0 = \frac{\pi^2k_B^2}{3} N(\epsilon_F)
\end{eqnarray}

where $k_B$ is the Boltzmann constant and $\lambda_{e-ph}$ is the electron-phonon coupling constant which can be considered negligibly small. From Eqs. 1 and 2, we have calculated $N(\epsilon_F)$ to be 13.6 states eV$^{-1}$ f.u$^{-1}$. This finite DOS is rather surprising for this material with strong insulating phase (discussed later). This estimated DOS is quite consistent with the calculated DOS values for Y$_2$Ir$_2$O$_7$ where major contribution comes from Ir 5$d$ atoms \cite{kalo}. It can be mentioned that finite $\gamma$ and resulting DOS has been observed for insulating pyrochlore iridate Eu$_2$Ir$_2$O$_7$ \cite{jun}. Given that this material is highly insulator, localized picture of DOS can explain this insulating behavior.  

The magnetic ordering as well as the structural symmetry have been theoretically discussed to play profound role in realizing various exotic electronic phases in the class of pyrochlore iridates. In case of materials lacking magnetic ordering, the recent calculation has predicted for various topologically insulated phases which comes out as an interplay between spin-orbit coupling effect and electronic correlation effect \cite{pesin}. The fact that long-range magnetic ordering breaks the time reversal symmetry which has motivated calculation to propose exotic Weyl-type semi-metal in Ir-based pyrochlores \cite{wan}. However, the structural symmetry also offers a vital clue for Weyl semi-metal as this phase demands breaking of either time reversal symmetry or inversion symmetry.

\begin{figure}
	\centering
		\includegraphics[width=8cm]{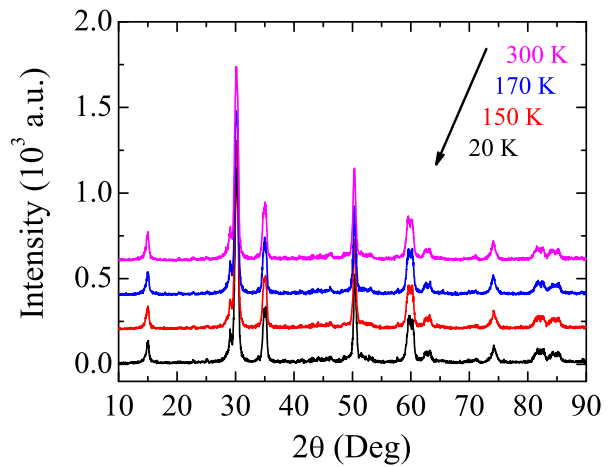}
	\caption{Representative XRD plots of Y$_2$Ir$_2$O$_7$ are shown at selective temperatures of 300, 170, 150 and 20 K. The plots at 170, 150 and 20 K are shifted vertically for clarity.}
	\label{fig:Fig5}
\end{figure}

In an aim to understand the evolution of structural symmetry across the magnetic irreversibility temperature $T_{irr}$ we have carried out temperature dependent XRD at various temperatures down to 20 K. Fig. 5 shows representative XRD plot of Y$_2$Ir$_2$O$_7$ taken at 300, 170, 150 and 20 K where the temperatures correspond to room temperature PM state (300 K), just above (170 K) and below (150 K) of $T_{irr}$ and low temperature magnetic state (20 K). It is clear in figure that XRD pattern does not exhibit any change with temperature in terms of peak splitting or arising of new peak(s). This primarily implies structural symmetry does not change with temperature or the spin ordering at $T_{irr}$ is not accompanied by any change in structural symmetry. All the XRD data collected at various temperatures are analyzed with Rietveld refinement program which shows the material retains cubic \textit{Fd$\bar{3}$m} symmetry down to lowest temperature 20 K. Though for the first time we present temperature dependent structural evolution using XRD, the similar structural behavior with temperature has also been previously shown from neutron diffraction for Y$_2$Ir$_2$O$_7$ \cite{shapiro}. Table I shows lattice parameters and crystallographic positions of all the atoms of Y$_2$Ir$_2$O$_7$ obtained from Rietveld analysis at two representative temperatures i.e., 300 and 20 K. It is observed that values do not change significantly over wide temperature range.

\begin{table}
\caption{\label{label} Unit cell parameters and crystallographic positions determined from the Rietveld profile refinement of the powder XRD patterns for Y$_2$Ir$_2$O$_7$ at 300 and 20 K. Here O and O$'$ refers to two different oxygen position (discussed in text).}
\begin{ruledtabular}
\begin{tabular}{ccc}
Parameters &300 K &20 K\\
          
          &\textit{Fd$\bar{3}$m} &\textit{Fd$\bar{3}$m}\\
\hline
a (\AA) &10.237(2) &10.219(2)\\
V (\AA${^3}$) &1072.7(3) &1067.2(3)\\
Y site &16d &16d\\
x &0.5 &0.5\\
y &0.5 &0.5\\
z &0.5 &0.5\\
Ir site &16c &16c\\
x &0.0 &0.0\\
y &0.0 &0.0\\
z &0.0 &0.0\\
O site &48f & 48f\\
x &0.336(3) &0.332(2)\\
y &0.125 &0.125\\
z &0.125 &0.125\\
O$'$ site &8b &8b\\
x &0.375 &0.375\\
y &0.375 &0.375\\
z &0.375 &0.375\\
\end{tabular}
\end{ruledtabular}
\end{table}      
From the Rietveld analysis of XRD data at different temperatures we have estimated unit cell parameter i.e., lattice constant $a$ which is shown in Fig. 6a as a function of temperature. With decreasing temperature, $a$ decreases but the change of $a(T)$ is not monotonic. Below $T_{irr}$ the $a$ decreases with faster rate while at lower temperature the rate slows down. The similar decrease of parameter $a$ with temperature has been observed from neutron diffraction data \cite{shapiro}. We have further plotted positional coordinate ($x$) of O atom which is related with IrO$_6$ octahedra as a function of temperature. Fig. 6b shows while there is some fluctuation in $x$ but the this parameter does not shows any consistent change with temperature which suggests IrO$_6$ octahedra are not systematically distorted with temperature. Using the lattice parameter and atomic positions, we have estimated the bond lengths ($d$) and bond angles ($<>$) between Y, Ir and O. Table II shows lengths and angles at representative two temperatures. The bond lengths $d_{Ir-O}$, $d_{Y-O}$ and $d_{Y-O'}$ are in agreement with general pyrochlore structure \cite{subram}. Out of four bond angels in basal plane, the opposite ones are found to be equal.     
      
\begin{table}[b]
\caption{\label{label} Bond lengths and bond angles determined using structural parameters for Y$_2$Ir$_2$O$_7$ at 300 and 20 K.  Here O and O$'$ refers to two different oxygen position (discussed in text).}
\begin{ruledtabular}
\begin{tabular}{ccc}
Parameters &300 K &20 K\\
          
          &\textit{Fd$\bar{3}$m} &\textit{Fd$\bar{3}$m}\\
\hline
$d_{Ir-O}$ (\AA) &2.0169(2) &1.9914(3)\\
$d_{Y-O}$ (\AA) &2.4615(2) &2.4921(1)\\
$d_{Y-O'}$ (\AA) &2.2164(2) &2.2125(2)\\
$<O-Ir-O>$ (deg) &80.9(1) &82.6(1)\\
$<O-Ir-O>$ (deg) &99.1(1) &97.4(1)\\
\end{tabular}
\end{ruledtabular}
\end{table}              
      
\begin{figure}
	\centering
		\includegraphics[width=7cm]{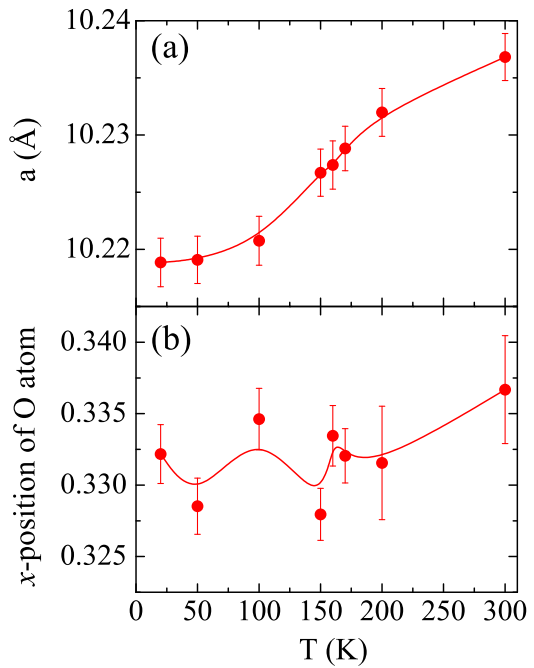}
	\caption{((a) Temperature variation in lattice parameters $a$ and (b) \textit{x}-position of O atom are shown for Y$_2$Ir$_2$O$_7$. Lines are guide to eyes.}
	\label{fig:Fig6}
\end{figure}

Above discussions imply that the material under study lacks long-range magnetic ordering while retains crystallographic symmetry down to low temperature. Note, that recent neutron diffraction measurement also do not find sign of long-range magnetic ordering at low temperature \cite{shapiro}. The magnetic irreversibility between M$_{ZFC}$ and M$_{FC}$ where a bifurcation arises at temperature $T_{irr}$ (Fig. 2) is, however, quite intriguing. The inherent frustration is quite prevalent in pyrochlore oxides (discussed above) and, indeed, we have found a reasonable value of frustration parameter `$f$' (2.4) for present Y$_2$Ir$_2$O$_7$. It is quite possible that the low temperature magnetic state in Y$_2$Ir$_2$O$_7$ exhibits glass-like behavior arising from Ir$^{4+}$ spin freezing. The characteristic property of glassy behavior is magnetic relaxation where the magnetic moment relaxes/evolves with time even at constant temperature and magnetic field unlike typical ferromagnetic or paramagnetic systems where the moment exhibits almost constant value in temperature-field phase diagram \cite{mydosh}. Moreover, aging behavior where the magnetic moment shows distinct value based on their history of application of magnetic field at particular temperature is another interesting property of glassy dynamics. We have examined the magnetic relaxation and aging behavior of our material. For the relaxation measurement, the material is cooled in zero field from 200 K ($>T_{irr}$) to 10 K. As temperature stabilizes, magnetic field of 1000 Oe is applied after a wait time $t_w$ and subsequently moment is measured as a function of time ($t$) for about 7200 s. The $t_w$ basically implies the time elapsed between stabilization of temperature and application of magnetic field. In our experiment we have used $t_w$ = 10$^{2}$, 10$^{3}$ and 10$^{4}$ s. For all values of $t_w$, we find that moment relaxes though relaxation rate varies. Also, we find the $M(0)$, the value of $M(t)$ at $t$ = 0 (after magnetic field is just switched on), increases with $t_w$ (not shown) which shows system ages during $t_w$ and tries to achieve low energy state giving different $M(0)$ though measured at same temperature and magnetic field. 

\begin{figure}
	\centering
		\includegraphics[width=8cm]{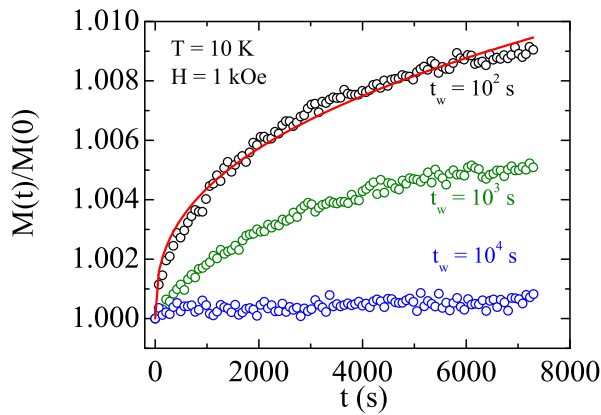}
	\caption{Normalized magnetic moment as a function of time is shown for different wait time $t_w$ for Y$_2$Ir$_2$O$_7$. The solid line shows representative fitting of data using Eq. 3.}
	\label{fig:Fig7}
\end{figure}

In Fig. 7, we show time evolution of magnetic moment normalized by M(0) for different $t_w$. As evident, for low value of $t_w$ = 10$^2$ s the moment changes significantly with no sign of saturation even after 2 h. This change in moment is about 0.9\% of initial value. Note, that residual ingredient materials i.e., Y$_2$O$_3$ and IrO$_2$ do not contribute in this magnetic relaxation as they are diamagnetic and paramagnetic, respectively in nature. With increasing $t_w$, the relative relaxation decreases. This time evolution of $M(t)$ has been explained with following functional form which is commonly known as stretched exponential relaxation \cite{chamberlin},

\begin{eqnarray}
	M(t) = M(0)\exp \left(\frac{t}{\tau}\right)^\beta
\end{eqnarray}
 
\begin{table}[b]
\caption{\label{label} Characteristic relaxation time $\tau$ and the exponent $\beta$ obtained using Eq. 3 are shown for different waiting time $t_w$ for Y$_2$Ir$_2$O$_7$.}
\begin{ruledtabular}
\begin{tabular}{ccc}
$t_w (s)$ &$\tau$ (s) &$\beta$\\      
\hline
10$^2$  &5.60 $\times$ 10$^9$ &0.33(1)\\
10$^3$  &1.78 $\times$ 10$^9$ &0.41(2)\\
10$^4$  &1.02 $\times$ 10$^8$ &0.82(8)\\
\end{tabular}
\end{ruledtabular}
\end{table} 

where $\tau$ is the characteristic relaxation time and the $\beta$ is the stretching exponent with value in the range of 0 $< \beta <$ 1. The $\beta$ = 1 recoves usual exponential relaxation behavior where the single energy barrier determines the relaxation behavior. The stretched exponential behavior, on other hand, originates if the system has multiple energy barrier which renders distribution of relaxation times. The solid line in Fig. 7 shows a representative fitting of our data using Eq. 3 for $t_w$ = 10$^2$ s. The parameters ($\tau$ and $\beta$) obtained from fitting are shown in Table III. The value of $\tau$ turns out very high. Both these characteristic parameters ($\tau$ and $\beta$) agree well with the values for glassy systems. At low temperatures, for instance, the typical values of $\tau$ and $\beta$ are obtained in the range of 10$^8$ - 10$^{10}$ s and 0.3 - 0.4, respectively for classical spin-glass system i.e., Ag:Mn \cite{hooger}. While the $\tau$ decreases with wait time the $\beta$, on contrary, increases. This magnetic relaxation as well as aging behavior in this material is quite interesting as it primarily suggests that the low temperature ground state in Y$_2$Ir$_2$O$_7$ is not an equilibrium state and the system relaxes to low energy state unlike long-range type magnetically ordered system. The decrease of $\tau$ and simultaneous increase of $\beta$ with $t_w$ imply that while aging the system tries to achieve low energy states and multiple energy barriers are gradually removed. The observed behavior akin to glassy systems in present material is quite intriguing. The calculation taking only classical Heisenberg interaction in pyrochlore lattice with antiferromagnetic interaction indeed show inherent frustration does not favor long-range ordering \cite{reimers}. The inclusion of additional Dzyaloshinsky-Moriya (DM) interaction, however, show a transition to long-range ordered phase at low temperature \cite{elhajal}. The sizable spin-orbit coupling effect in iridium oxides may induce Dzyaloshinsky-Moriya type interaction but this can not be generalized as Pr$_2$Ir$_2$O$_7$ does not exhibit magnetically ordered phase down to low temperature \cite{nakatsuji}.    
     
\begin{figure}
	\centering
		\includegraphics[width=7cm]{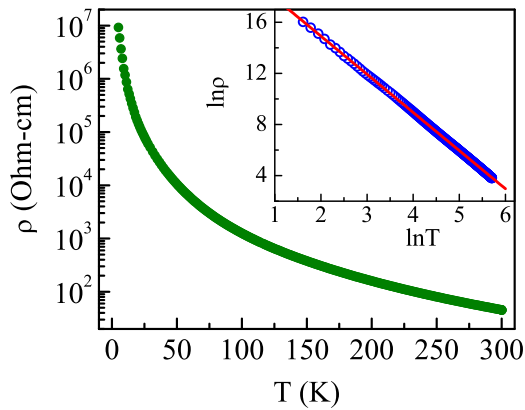}
	\caption{Temperature variation in electrical resistivity is shown for Y$_2$Ir$_2$O$_7$ in semilogarithmic plot. Inset shows same resistivity data in ln-ln plot. The solid line shows straight line fitting.}
	\label{fig:Fig8}
\end{figure}

Now we look at the electronic transport behavior in this material. An insulating behavior is evident for Y$_2$Ir$_2$O$_7$ in previous studies, however, the resistivity value at low temperature has shown wide variation \cite{disseler,soda,fukazawa1,zhu}. Fig. 8 shows resistivity ($\rho$) of this material as a function of temperature. An insulating behavior is observed throughout the temperature regime, even across $T_{irr}$. The figure shows resistivity increases by couple of orders at low temperatures. In contrast to thermally activated hopping transport which is commonly observed for insulating or semiconducting systems, we find that temperature dependence of resistivity follows a power law behavior as,

\begin{eqnarray}
	\rho = \rho_0 T^{-n}
\end{eqnarray}
    
Inset of Fig. 8 shows ln-ln plot of $\rho(T)$ data and the fitted straight line confirms the validity of Eq. 4. The straight line fitting yields exponent $n$ = 2.987(2). The similar power law driven electronic transport has also been observed for Y$_2$Ir$_2$O$_7$ with the exponent $n$ = 4 \cite{disseler}.
 
We have also looked into the electronic transport behavior in presence of magnetic field. The response of electrical resistance in application of magnetic field or the magnetoresistance (MR) effect, which is defined as $\Delta \rho/\rho(0)$ = $\left[\rho(H) - \rho(0)\right]/\rho(0)$, has attracted lot of interest for materials with strong spin-orbit coupling effect. Usually, a positive MR has been observed in such materials which is largely explained employing `weak antilocalization' effect. Exemplary compounds are  Bi$_2$Se$_3$ \cite{chen1}, Bi$_2$Te$_3$ \cite{he}, Au covered Mg film \cite{berg}, even in Ir-based Na$_2$IrO$_3$ films \cite{jender}. In fact, to our knowledge the present MR investigation is first of its kind in Y$_2$Ir$_2$O$_7$, even in family of pyrochlore iridates. The main panel of Fig. 9 shows observed MR behavior as a function of magnetic field. As evident, the resistivity decreases in magnetic field exhibiting negative MR behavior. Though at maximum available field of 8 Tesla the MR continues to increase without sign of saturation, we find MR of value $\sim$ 11\% at this field.

\begin{figure}
	\centering
		\includegraphics[width=8cm]{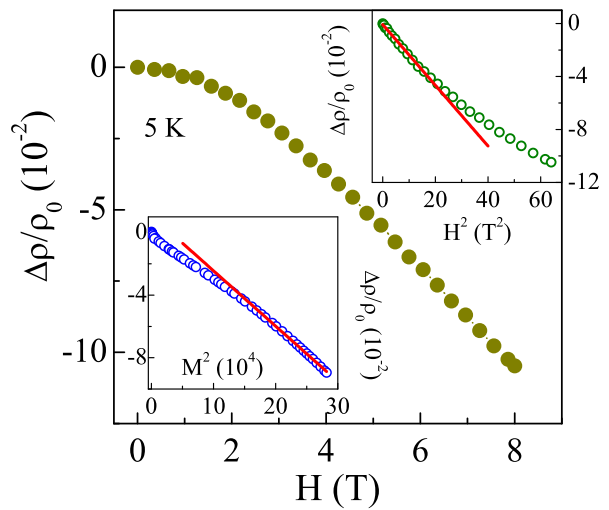}
		\caption{Magnetoresistance as a function of field has been shown for Y$_2$Ir$_2$O$_7$. Upper inset shows quadratic field dependence of MR and lower inset shows linear variation of MR with square of magnetization (emu/mol) for Y$_2$Ir$_2$O$_7$.}
	\label{fig:Fig9}
\end{figure}

From theoretical understanding, the negative MR at low temperatures in weakly disordered systems has been largely explained using `weak localization' effect. The quantum interference (QI) of electronic wave functions is the basis of this localization effect and this is treated as quantum correction to the classical Drude equation for conductivity. In this scenario, application of magnetic field tend to destroy the QI effect, hence promotes the conductivity which results in negative MR. Sivan, Entin-Wohlman and Imry\cite{imry} applied a method of critical path analysis to understand the conductivity over many random paths and found a quadratic field dependence of MR. In fact, this behavior has been experimentally observed in variety kind of systems. The upper inset of Fig. 9 shows MR at 5 K varies linearly with $H^2$ up to field $\sim$ 4.3 Tesla. This quadratic field dependence of MR implies in low field regime negative MR originates from weak localization effect. Further, what we find interesting is that above this field (4.3 Tesla), MR is found to vary linearly with $M^2$ (lower inset of Fig. 9). Such correlation between electrical resistivity and magnetization is evident for pyrochlore iridates A$_2$Ir$_2$O$_7$ with A = Pr and Nd ascribing it to Kondo-like screening behavior \cite{nakatsuji,disseler1}, but this behavior is observed in present Y$_2$Ir$_2$O$_7$ in higher field regime. More surprisingly, unlike the previous Pr and Nd based compounds, in present material with Y$^{3+}$ ion sitting at A-site do not posses any $f$ electrons. We speculate that internal fields originating from more orderly spins in higher fields is responsible for this quadratic dependence of MR with magnetization. Nonetheless, electronic transport behavior in magnetic field in iridate systems needs to be understood thoroughly by employing theoretical calculations as well as studying other similar systems.        
             
\section{Conclusion}
In conclusion, detailed structural, magnetic, thermodynamic and transport properties are studied in polycrystalline sample of pyrochlore iridate Y$_2$Ir$_2$O$_7$. The material shows magnetic irreversibility at temperature $T_{irr}$ $\sim$ 160 K. However, below $T_{irr}$ no sign of long-range type ferromagnetic ordering is observed. The absence of anomaly in specific heat around $T_{irr}$ also favors absence of long-range type magnetic ordering. The specific heat data indicate reasonable DOS at Fermi level. Temperature dependent XRD data shows the material under study retains its structural symmetry down to low temperature. From magnetic relaxation and aging behavior we infer that Y$_2$Ir$_2$O$_7$ shows glassy behavior at low temperature. The system is insulator throughout the temperature regime where the electronic transport mechanism follows power law behavior with temperature. The insulating behavior suggests perhaps DOS are localized. The negative magnetoresistance in this SOC dominated material is quite intriguing which is explained with quantum interference effect.   

\section{Acknowledgment}  
We are thankful to AIRF, JNU for low temperature XRD measurements and Manoj Pratap Singh for the help in measurements. We acknowledge UGC-DAE CSR, Indore for magnetization and resistivity measurements. We sincerely thank Alok Banerjee and Rajeev Rawat for the magnetization, specific heat and resistivity data and fruitful discussions. We thank Kranti Kumar and Sachin Kumar for the helps in measurements. HK acknowledges UGC, India for BSR fellowship.

\end{document}